\documentclass[1, 12pt]{elsarticle}
\usepackage[english]{babel}
\usepackage{amsmath}
\usepackage[toc,page]{appendix}
\usepackage{amssymb}       
\usepackage{amsthm}       
\usepackage{natbib}        
\usepackage{graphicx}
\usepackage{booktabs}
\usepackage{subfigure}      
\usepackage{color}
\usepackage{gensymb} 

\journal{Acta Materialia, {\rm accepted for publication}}

\newcommand{\tensor}[1]{\boldsymbol{#1}}
\newcommand{\ftensor}[1]{\mathbb{#1}}

\newcommand{\rd}{\mathrm{d}}

\begin{document}

\begin{frontmatter}

\title{Multiscale modelling of precipitation hardening in Al-Cu alloys: dislocation dynamics simulations and experimental validation} 

\author{R. Santos-G{\"u}emes$^{1, 2}$}
\author{B. Bell{\'o}n$^{1, 2}$}
\author{G. Esteban-Manzanares$^{1, 2}$}
\author{J. Segurado$^{1, 2}$}
\author{L. Capolungo$^{3}$}
\author{J. LLorca$^{1, 2, }$\corref{cor1}}
\address{$^1$ IMDEA Materials Institute, C/ Eric Kandel 2, 28906, Getafe, Madrid, Spain. \\  $^2$ Department of Materials Science, Polytechnic University of Madrid/Universidad Polit\'ecnica de Madrid, E. T. S. de Ingenieros de Caminos. 28040 - Madrid, Spain. \\  $^3$ Material Science and Technology Division, MST-8, Los Alamos National Laboratory, Los Alamos 87545 NM, USA.}

\cortext[cor1]{Corresponding author}

\begin{abstract}

The mechanisms of dislocation/precipitate interactions were analyzed in an Al-Cu alloy containing a homogeneous dispersion of $\theta'$ precipitates by means of discrete dislocation dynamics simulations. The simulations were carried out within the framework of the discrete-continuous method and the precipitates were assumed to be impenetrable by dislocations. The main parameters that determine the dislocation/precipitate interactions (elastic mismatch, stress-free transformation strains, dislocation mobility and cross-slip rate) were obtained from atomistic simulations, while the size, shape, spatial distribution and volume fraction of the precipitates were obtained from transmission electron microscopy. The predictions of the critical resolved shear stress (including the contribution of solid solution) were in agreement with the experimental results obtained by means of compression tests in micropillars of the Al-Cu alloy oriented for single slip. The simulations revealed  that the most important contribution to the precipitation hardening of the alloy was provided by the stress-free transformation strains followed by the solution hardening and the Orowan mechanism due to the bow-out of the dislocations around the precipitates.
\end{abstract}

\begin{keyword}
Dislocation dynamics, precipitate strengthening, multiscale modeling, Al-Cu alloys
\end{keyword}

\end{frontmatter}

\section{Introduction}
\label{sec:intro}

Precipitation hardening is one of the most effective mechanisms for increasing the strength of metallic alloys \citep{ardell1985precipitation,BalakrishnaBhat1980,Martin1998_PrecipitationHardening}. Dislocation glide along the crystallographic planes is hindered by the presence of nm-sized precipitates, increasing the critical resolved shear stress (CRSS) to move dislocations. The efficiency of precipitation hardening is known to depend on geometrical factors (size, shape, volume fraction and spatial distribution of the precipitates with respect to the glide plane)  and on the actual mechanisms of dislocation/precipitate interactions. In general, small precipitates ($<$ 10 nm) coherent with the matrix tend to be sheared by dislocations while large precipitates ($>$ 50 nm) with incoherent interfaces are by-passed by the formation of Orowan loops or by dislocation cross-slip to other crystallographic planes.

Modelling precipitate hardening in the case of obstacles impenetrable to dislocations was pioneered by Orowan using a constant line tension model and assuming that the precipitates were arranged in a square lattice in the slip plane and that  the precipitate cross-section in the slip plane was circular \citep{Orowan1948}.  He predicted that the dislocation bowed around the obstacle leaving a dislocation loop around the precipitate and that the CRSS was inversely proportional to the precipitate spacing. These results were in good qualitative agreement with experimental observations and the Orowan model has been refined over the years leading to analytical expressions that  take into account the influence of the interaction stresses between the dislocation segments during the formation of the loop \citep{Bacon1973_BKS}, the spatial distribution of the precipitates in the glide plane \citep{Foreman1966_FM, Kocks1966} as well as the precipitate geometry and crystallographic orientation \citep{nie1996effect} on the CRSS. Nevertheless, these simple models could not provide quantitative estimations nor to explain why precipitation hardening is very efficient in some metallic alloys (such as Al-Cu and Ni-Al) but not in others (Mg alloys) \citep{N14}.

More detailed analyses of the dislocation/precipitate interactions were carried within the framework of discrete dislocation dynamics (DDD) simulations. For instance, Xiang {\it et al.} \cite{Xiang2004_LevelSet, Xiang2006_LevelSet}  used the level-set method to study the interaction between dislocations and spherical precipitates and included the effect of the misfit dilatational strain between the matrix and the precipitate for both shearable and non-shearable precipitates. They reported a large variety of by-pass mechanisms, including paths not restricted to the initial dislocation glide plane, which point out the importance of dislocation cross-slip or climb (which are thermally activated processes) in precipitation hardening. It should be noted that these simulations did not take into account the crystallography of slip, leading to limitations in the precise modelling of the out-of-plane dislocation mobility.

Further investigations analyzed the interaction of dislocations with spherical precipitates taking into account the crystallography of slip \citep{Monnet2006_DDD_precs, Monnet2011_DDD_precs, Queyreau2010_superposition} and and of dislocation cross-slip \citep{Shin2003_DDD_FEM, Monnet2006_DDD_precs} as well as the effect of the image stresses induced by the elastic modulus mismatch between the matrix and the spherical precipitate  \citep{Takahashi2008_DDD_BEM, Takahashi2011_DDD_FEM, Shin2003_DDD_FEM}. The results of these simulations were compared with the predictions of the Orowan-based models, indicating the limitations of the analytical approximations \citep{Monnet2006_DDD_precs}. Other authors tried to account for the  details of the dislocation/precipitate interactions from atomistic simulations \citep{monnet2015multiscale, lehtinen2016multiscale} and the dislocation mobility and core energy were also obtained from calculations at lower length scales in \citep{lehtinen2016multiscale}, reinforcing the multiscale approach. Nevertheless, all these simulations were still far away of providing parameter-free, quantitative estimations of precipitate-strengthening in metallic alloys because  they did not considered very important factors such as the actual shape and crystallographic orientation of the precipitates and of the coherency or transformation stresses around the precipitates. More recent simulations \citep{GFM15, ZSD17, Santos-Guemes2018} have demonstrated the large contribution of these factors to the CRSS but direct comparison with experimental data including all the relevant physical processes that determine the dislocation/precipitate interactions are not available.

The objective of this investigation is to demonstrate that the CRSS due to precipitation-hardening can be quantitatively predicted from DDD simulations by taking into account all the factors that control the dislocation/precipitate interactions. They include the dislocation mobility and cross-slip probability, the interaction stresses between the dislocation and the precipitate due to the elastic mismatch and to the transformation strains around the precipitate, the solid-solution hardening contribution (which is always present in precipitation-hardening alloys) as well as the size, shape and spatial distribution of the precipitates in the matrix. Moreover, all the parameters in the DDD simulations were obtained from {\it ab initio} and atomistic simulations or experimental observations without any adjustable parameters. The investigation was focussed in an Al-Cu alloy strengthened with $\theta'$ precipitates and the predictions obtained by means of the DDD simulations were validated against the experimental results in this alloy. The simulations showed the contribution of the different mechanisms (Orowan looping, elastic mismatch, transformation strains, solution hardening, dislocation cross-slip) to the overall hardening and explained why Al-Cu alloys are efficiently strengthened by $\theta'$ precipitates. Moreover, they also indicate the key parameters that should be taking into account to design new alloys with optimum response to precipitation hardening.

The paper is organized as follows. After the introduction, the experimental results of the precipitate distribution and of the CRSS are presented in section \ref{sec:Mat_and_exp}. The matrix and precipitate properties that influence the dislocation/precipitate interactions (elastic constants, transformation strains) obtained from {\it ab initio} simulations are reported in section \ref{sec:MP}. The DDD strategy, including the details of the dislocation mobility and cross-slip laws, are shown in section \ref{sec:Methodology} while the results of the DDD simulations and the experimental validation can be found in section \ref{sec:Results}. The conclusions of the paper are summarized in section \ref{sec:Conclusions}.

\section{Material and experimental results}
\label{sec:Mat_and_exp}

\begin{figure}
\centering
\includegraphics[width=\textwidth]{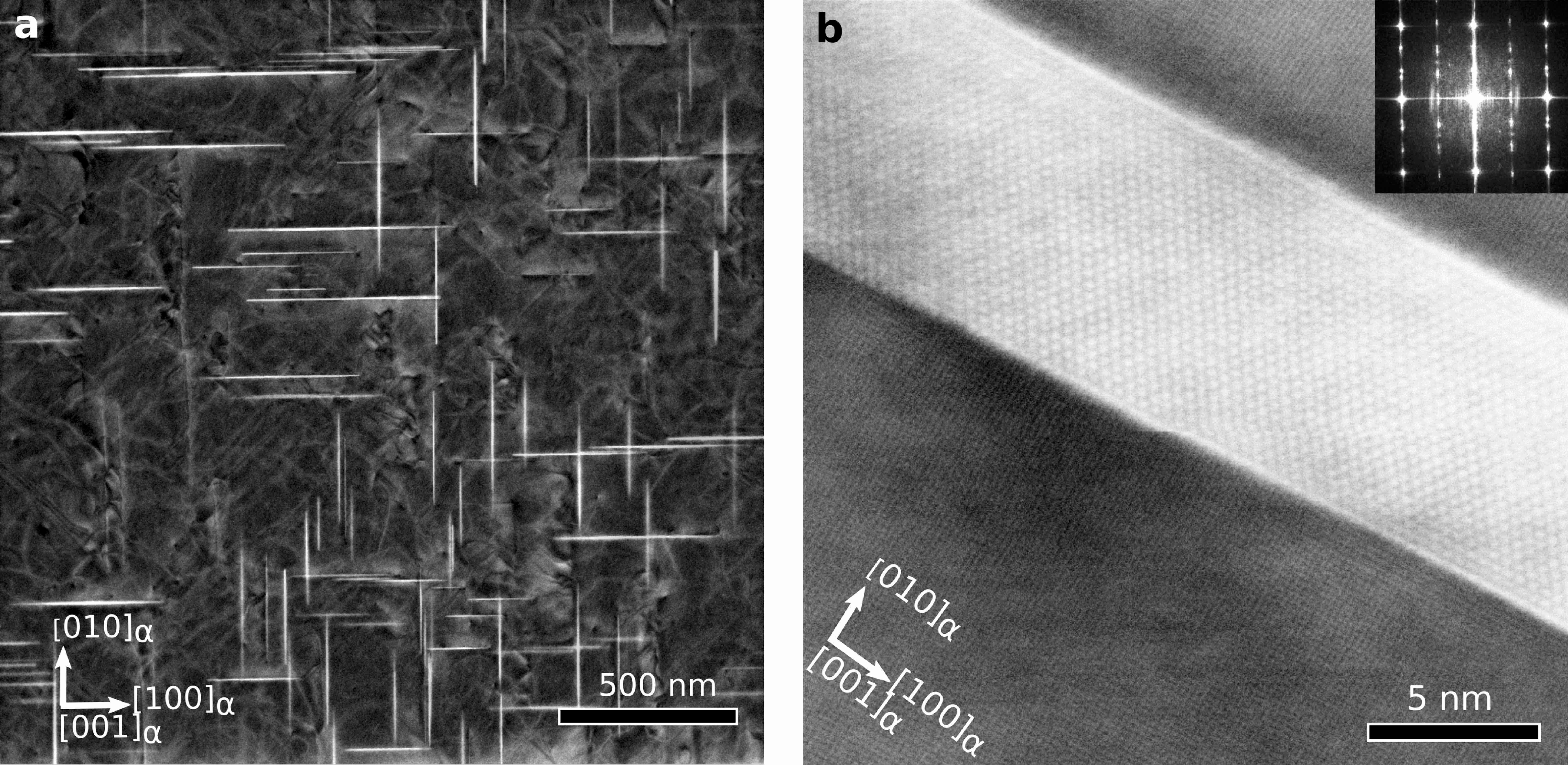}
\caption{HAADF micrographs of the Al-4\%Cu specimens aged at 180$^{\circ}$C for 168 hours. The FFT of the $\theta'$ precipitates is shown in the inset of (b). The electron beam was close to the $<$001$>_\alpha$ orientation.}  
\label{TEM}
\end{figure}

An Al - 4\% wt. Cu (1.7 at.\%) alloy was prepared using high-purity metals by casting in an induction furnace (VSG 002 DS, PVA TePla). Samples were machined from the ingot and subjected to homogenization and solution heat treatments during 22 h at $540^{\circ}$C followed by quenching in water, leading to polycrystals with large average grain size (above several hundreds $\mu$m). Afterwards, the samples were aged at 180$^{\circ}$C for 168 hours. The structure, size, shape and spatial distribution of the precipitates was carefully characterized by means of transmission electron microscopy (TEM) (FEI Talos) \citep{Rodriguez-Veiga2018}. The microstructure of the aged alloys was made up of a homogeneous distribution of  precipitates in the Al matrix, as shown in the high-angle annular dark-field mode  (HAADF) micrographs in Fig. \ref{TEM}. The precipitates were identified as $\theta'$ (Al$_2$Cu) from the reciprocal lattice images  obtained using high resolution dark field TEM images along with fast Fourier transform (FFT) of the individual precipitates (Fig. \ref{TEM}b). The precipitates had a disk shape and were parallel to the \{100\} planes of the FCC  $\alpha$-Al matrix, leading to three different orientation variants, of which two are visible in the TEM images with the electron beam parallel to  $<$100$>_\alpha$.  The precipitate volume fraction, diameter and thickness were carefully measured in \citep{Rodriguez-Veiga2018} from the TEM micrographs using the methodology presented in \cite{NM08}.  The precipitate diameters presented a log-normal distribution and the average values of the precipitate volume fraction, diameter and thickness are shown in Table \ref{tab:precipitate-size}. 

\begin{table}[h]
	\centering
	\begin{tabular}{ccccccc}	
		\toprule
		$f$ (\%) & $D$ (nm) & $t$ (nm) & $AR$ & \\
		\midrule
		  1.0 $ \pm $ 0.5 & 342 $\pm$ 47 & 9.0 $ \pm $ 0.6 & 39 $ \pm $ 6.6 & \\
		\bottomrule
	\end{tabular}%
		\caption{Volume fraction ($f$), diameter ($D$), thickness ($t$) and aspect ratio ($AR$) of the $ \theta' $ precipitates in the Al-4\%Cu specimens aged at 180$^{\circ}$C for 168 h. \citep{Rodriguez-Veiga2018}.}
	\label{tab:precipitate-size}%
\end{table}%

Micropillars with an average aspect ratio in the range 2:1 to 3:1 were milled using a focus ion beam (FEI Helios NanoLab 600i) from the polycrystal in grains oriented single slip (the micropillar axis was close the $<$123$>$ direction)
using the methodology detailed in \citep{WLA19}. Micropillars of circular cross-section and different diameters  (from 1 to 7 $\mu$m) as well as of square cross-section (5 $\times$ 5 $\mu$m$^2$) were carved to measure the CRSS in compression. The taper angle was small ($<$ 2$^\circ$) for circular micropillars and negligible ($<$ 1$^\circ$) for square ones. Compression of the micropillars was carried out using a circular diamond flat punch of 10 $\mu$m of diameter using a nanoindentor (Hysitron TI950). Tests were carried out in displacement control at an average strain rate of $ \approx 10^{-3}$ s$^{-1}$ up to 10\% strain.
Circular micropillars with different diameter were deformed using a high-load transducer (up to 900 mN) that can deform the micropillar with the largest diameter. Square micropillars were deformed with a low-load transducer (12 mN) that was able to capture the strain bursts.

 The engineering stress-strain curves were calculated from the applied load using the upper cross-sectional area and length of the pillars, measured after the milling process. The engineering stress was transformed into the resolved shear stress using the Schmid factor of the most suitable oriented slip system in the micropillar. The CRSS - which correspond to the beginning of plastic deformation when the dislocations are able to by-pass the precipitates- was determined  from the resolved shear stress-strain curves at a plastic strain of 0.02$\%$ \citep{Proudhon2008,ASMHandbookV2}.

\begin{figure}[t!]
\centering
\includegraphics[width=0.55\textwidth]{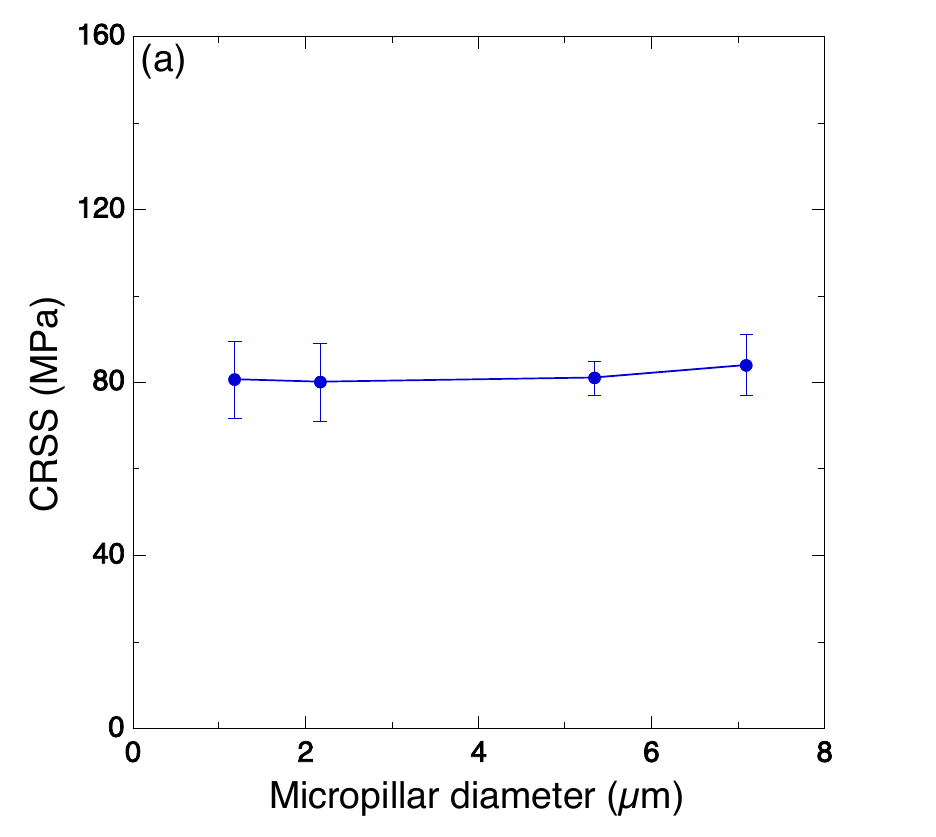}
\includegraphics[width=\textwidth]{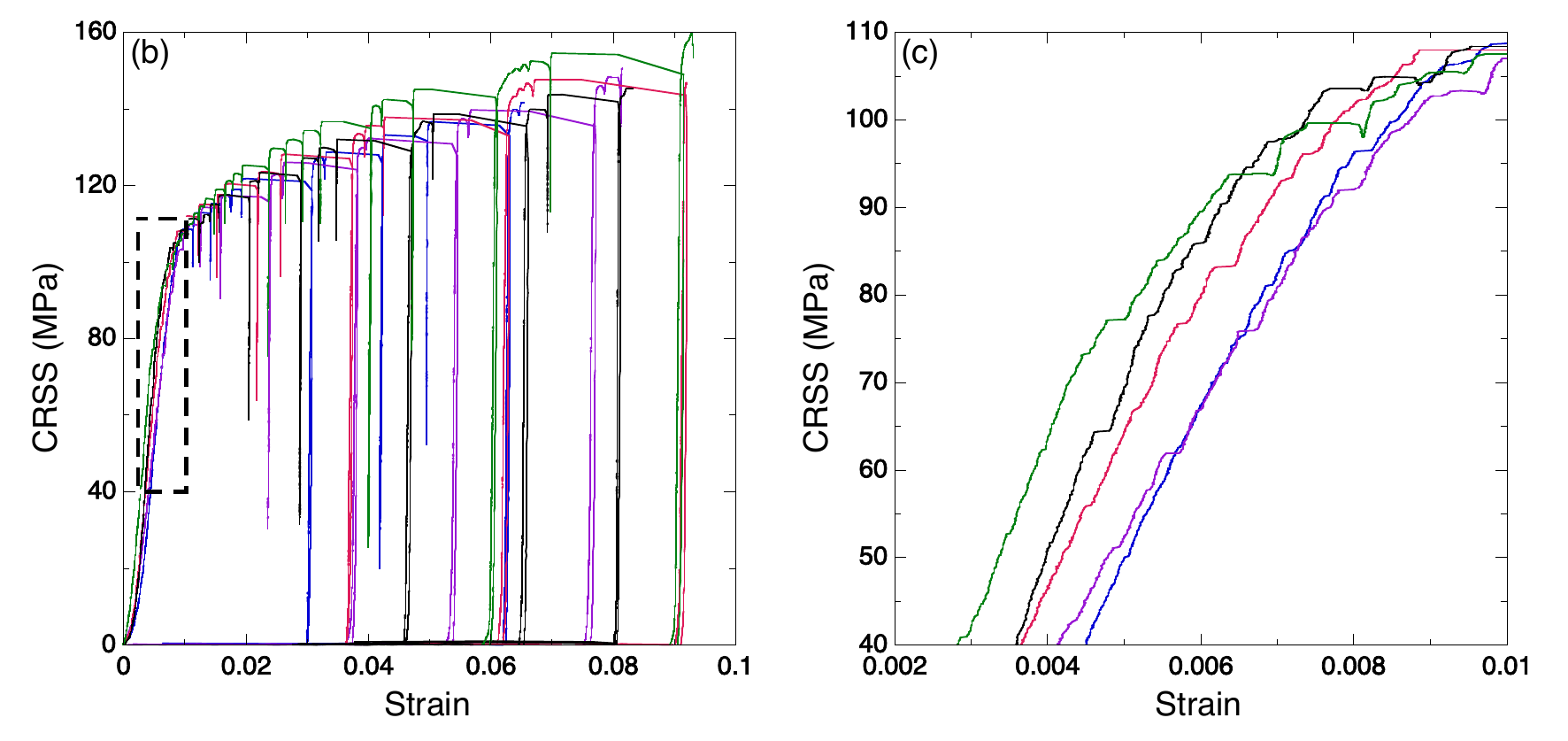}
\caption{(a) Effect of micropillar diameter on the CRSS of the Al-4\% Cu alloy aged at 180$^{\circ}$C for 168 hours. (b) Resolved shear stress-strain curves of square micropillars of  5 $\times$ 5 $\mu$m$^2$ oriented for single slip. (c) Detail of region marked with a rectangle in (b) showing the small strain bursts associated with the onset of dislocation plasticity.}  
\label{Micropillar_tests}
\end{figure}

The average CRSS measured in the circular micropillars of different diameter is plotted in Fig. \ref{Micropillar_tests}a together with the standard deviation corresponding to three tests for each diameter. These results show that the CRSS was independent of the micropillar diameter and, thus, representative of the CRSS of the bulk. Similar behavior was found in the case of IN718 superalloy \citep{CGJ15} because the critical length scale that controls the CRSS in these precipitation-hardened alloys is the distance between precipitates, which is much smaller than the micropillar diameter. 
The resolved shear stress-strain curves of five micropillars of square cross-section are plotted in Fig. \ref{Micropillar_tests}b. The experimental scatter was very small and the region corresponding to the onset of plastic deformation is detailed in Fig. \ref{Micropillar_tests}c. Small strain bursts were observed when the resolved shear stress was around 80 MPa, which is in agreement with the value of  80 $\pm$ 6 MPa, determined from the 0.02\% plastic strain criterion.

\section{Matrix and precipitate properties}
\label{sec:MP}

\begin{figure}
\centering
\includegraphics[width=0.8\textwidth]{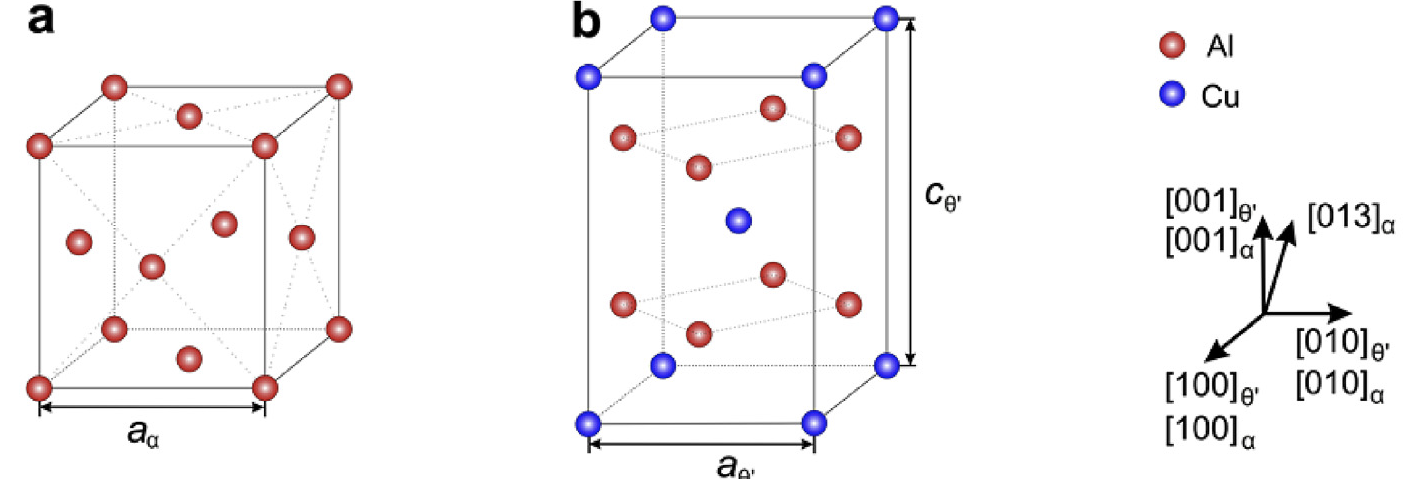}
\caption{Unit cells of (a) $\alpha$-Al  matrix and (b) $\theta'$ precipitates. }  
\label{UC}
\end{figure}

The unit cells of the $\alpha$-Al matrix and of the $\theta'$ precipitate are shown in Fig. \ref{UC}. The $\alpha$-Al matrix has a face-centered cubic (FCC) structure with a lattice parameter $a_{\alpha}$ = 0.405 nm. The $\theta'$ precipitate has a body-centered tetragonal (BCT) structure (space group $I4/mmm$) with lattice parameters $a_{\theta'}$ = 0.404 nm and $c_{\theta'}$ = 0.580 nm) \citep{N14}.  The orientation relationship between $\theta'$ and $\alpha$-Al is (001)$_\theta' \parallel (001)_\alpha$ and [100]$_\theta' \parallel [100]_\alpha$, leading to three orientation variants for the $\theta'$ precipitates \citep{LBL17}. The $\theta'$ precipitates have a disk-shape and the broad faces of the disk are nearly fully coherent with the $\alpha$-Al matrix while the edges of the plates are semi-coherent.

The phase transformation from the FCC $\alpha$  lattice to the BCT $\theta'$ lattice has been carefully analyzed \citep{DW83, NM99} and involves a  homogeneous shear of the whole cell by an angle arctan(1/3) around the coincident 
[100]$_\alpha \parallel [100]_\theta'$ orientations,  leading to a lattice correspondence $[013]_\alpha \rightarrow [001]_\theta'$ and $[010]_\alpha \rightarrow [010]_\theta'$ (Fig. \ref{UC}). The transformation matrix, $\mathbf{T}$, that links the lattice parameters of $\alpha$-Al ($\mathbf{e}_\alpha$) and $\theta'$ ($\mathbf{e}_{\theta'}$)   through $\mathbf{T} \mathbf{e}_\alpha = \mathbf{e}_{\theta'}$  is given by \citep{GLS12, LBL17}:

\begin{equation}
\begin{array}{ccc}
\mathbf{T}=\begin{pmatrix}
    a_{\theta'}/a_\alpha &0&0\\  0&a_{\theta'}/a_\alpha&-1/3\\ 0&0&c_{\theta'}/1.5a_\alpha
  \end{pmatrix}
\end{array}
\label{eq:T}
\end{equation}

\noindent for the precipitate variant in Fig. \ref{UC}. The transformation matrix includes both strains and rigid body rotations and the corresponding Lagrangian stress-free transformation strain (SFTS), $\boldsymbol{\epsilon}^0$, can be expressed in the axis defined by the FCC Al as \citep{LBL17}

\begin{equation}
\boldsymbol{\epsilon}^0 = \frac{1}{2}({\mathbf T}^T {\mathbf T} - {\mathbf I})
\label{eq:SFTS}
\end{equation}

\noindent where ${\mathbf I}$ stands for the identity matrix. 

The elastic constants (as well as the lattice parameters) of  $\alpha$-Al and $\theta'$ were determined from first principles calculations in a previous investigation \citep{Santos-Guemes2018}. and are depicted in Tables \ref{Tab:ECA} and \ref{Tab:ECT}. It should be noted that the predictions of the elastic constants of $\alpha$-Al are very close to the experimental results \citep{SS02}.

\begin{center}
\begin{table}[h]
\begin{center}
\begin{tabular}{cccc}
\toprule
  $C_{11}$ & $C_{12}$ & $C_{44}$\\
\midrule
 110.4  & 60.0 & 31.6 \\
\bottomrule
\end{tabular}
\end{center}
\caption{Elastic constants (in GPa) of $\alpha$ - Al obtained from first principles calculations \citep{Santos-Guemes2018}.}
\label{Tab:ECA}
\end{table}
\end{center}

\begin{center}
\begin{table}[h]
\begin{center}
\begin{tabular}{cccccc}
\toprule
 $C_{11}=C_{22}$ & $C_{12}$ & $C_{13}=C_{23}$ & $C_{33}$ & $C_{44}=C_{55}$ & $C_{66}$\\
\midrule
 212.6 & 39.9 & 61.4 & 173.5 & 82.8 & 44.8\\
\bottomrule
\end{tabular}
\end{center}
\caption{Elastic constants (in GPa) of  $\theta'$ - Al$_{2}$Cu precipitates from first principles calculations \citep{Santos-Guemes2018}.}
\label{Tab:ECT}
\end{table}
\end{center}

\section{Dislocation dynamics simulations}
\label{sec:Methodology}

Simulation of the dislocation/precipitate interactions within the framework of DDD was carried using the discrete-continuous approach, originally developed by Lemarchand {\it et al.} \cite{Lemarchand2001_DCM}. In this strategy, dislocation loops are represented by plate-like inclusions with an eigenstrain that represents the plastic strain associated with the area sheared by the dislocation. The dislocation loop is discretized in segments and the displacement of each segment in each simulation step depends on the Peach-Koehler force. The plastic strain is computed directly from the area sheared by the dislocation loop and this information is used to determine the mechanical fields in the simulation domain. The main advantages of this simulation strategy is that the computational efforts do not increase with the square of the dislocation segments in the simulation and, more importantly, that it is straightforward to introduce elastic heterogeneities, anisotropic materials and transformation strains (as well as other eigenstrains)  in the simulation without increasing the computational effort. However, this strategy requires a very fine discretization of the whole simulation domain and the solution of the mechanical fields using the finite element method hindered the applicability of this approach. This limitation was overcome with the use of a solver based on the FFT, leading to an efficient DDD tool for problems involving elastic and plastic heterogeneities as well as eigenstrains \citep{Bertin2018}. This strategy was recently modified by the authors to account for the interaction of a  single straight dislocation segment (with either edge or screw character) with one precipitate \citep{Santos-Guemes2018} and it is extended here to a domain containing a homogeneous dispersion of precipitates.

The basic elements of the DDD strategy are reviewed below for the sake of completion and more details can be found in  \cite{Bertin2018, Santos-Guemes2018}. In addition, the multiscale nature of the approach to model precipitation-hardening is emphasized: the parameters that control the dislocation mobility as well as the interaction of dislocations with the precipitates were obtained from simulations at lower length scales, leading to predictions that are free from adjustable parameters.

\subsection{Dislocation mobility}
 Dislocations are discretized into segments limited by nodes. The velocity of node $i$ in the glide plane, $\mathbf{v}_i$,  is given by

\begin{equation} \label{eq:mobility}
\mathbf{v}_i = \left\{
 \begin{array}{ll}
 [\mathbf{F}_i^{g}-F_i^{ss}(\mathbf{F}_i^{g}/|\mathbf{F}_i^{g}|)]/B & \mathrm{if} \quad |\mathbf{F}_i^{g}| > F_i^{ss}
\\ \\
 0 &  \mathrm{if} \quad|\mathbf{F}_i^{g}| \leq F_i^{ss}
 \end{array}
 \right.
\end{equation}

\noindent where $\mathbf{F}_i^{g}$ is the projection of the nodal force, $\mathbf{F}_i$, on the glide plane (characterized by the slip plane normal $\mathbf{n}$) according to

\begin{equation}
\mathbf{F}_{i}^{g} =  \mathbf{F}_{i} - (\mathbf{F}_{i} \cdot \mathbf{n})\mathbf{n} \label{4}
\end{equation}

\noindent $B$ is the viscous drag coefficient that depends on the dislocation character and $F_i^{ss}$ is 
 a force threshold due to solid solution strengthening whose effect is reducing the effective nodal glide forces, and thus, the corresponding nodal velocity.

The nodal force $\mathbf{F}_i$ is determined as

 \begin{equation}
\mathbf{F}_i = \sum_j \mathbf{f}_{ij}
\end{equation}

\noindent where $\mathbf{f}_{ij}$ is the force acting on the segment $ij$ (limited by nodes $i$ and $j$), which is computed according to

 \begin{equation}
\mathbf{f}_{ij} = \int_{\mathbf{x}_i}^{\mathbf{x}_j} N_i(\mathbf{x}) \mathbf{f}_{ij}^{pk}(\mathbf{x}) \rd \mathbf{x}
\end{equation}

\noindent where $N_i$ is the interpolation function associated to node $i$ and $\mathbf{f}_{ij}^{pk}$ is the Peach-Koehler force given by

 \begin{equation}
\mathbf{f}_{ij}^{pk}(\mathbf{x}) =  \Big(\boldsymbol{\sigma}(\mathbf{x}) \cdot  \mathbf{b}_{ij}\Big) \times \mathbf{\hat t}_{ij} \label{7}
\end{equation}

\noindent where $\mathbf{b}_{ij}$ is the Burgers vector of the segment $ij$ and $\mathbf{\hat t}_{ij}$ the unit vector parallel to the dislocation line.

The threshold nodal force in the glide plane due to solid solution, $F_i^{ss}$, can be determined from eqs. \eqref{4}  to  \eqref{7}  by assuming that the Peach-Koehler force accounting for solid solution strengthening in each dislocation segment, $f^{pk,ss}_{ij}$, is constant and equal to

\begin{equation}
f^{pk,ss}_{ij}=\tau_{ss}b
\end{equation}

\noindent where $\tau_{ss}$ is contribution of solid solution to the CRSS. This approach to take into account solid solution in DDD simulations has been used by other authors  \citep{Queyreau2010_superposition}.

The viscous drag coefficient $B$ in eq. \eqref{eq:mobility} was determined in Al by Cho {\it et al.} \citep{Molinari2017_Mobility} as a function of the dislocation character using molecular dynamics simulations and these results were approximated by an analytical function reported in \citep{Santos-Guemes2018}.

\subsection{Dislocation cross-slip}
\label{sec:CS}

Cross-slip of screw dislocation segments during the simulations was accounted for through an Arrhenius-type equation characteristic of thermally activated processes \citep{Kubin1992,Hussein2015}. In this investigation, the cross-slip probability $P$ of a screw dislocation segment of length $L$ in Al was determined  as a function of the applied stress $\boldsymbol\sigma$ and temperature $T$ from molecular dynamics simulations within harmonic transition state theory framework according to \citep{Esteban-Manzanares2019_CS} 

\begin{equation}
P(\boldsymbol\sigma, T) = \nu_{eff} \frac{L}{L_n} e ^{-\left[\Delta H(\boldsymbol\sigma)\left(1-\frac{T}{T_m}\right)\right]/k_bT} \Delta t
\label{eq:CS}
\end{equation}

\noindent where $k_b$ stands for the Boltzmann constant,  $\Delta t$ is the time step in the DDD simulation and $L_n$ (= 2.8 nm) the nucleation length, i.e. the length between constrictions to nucleate the cross-slip process, which  was determined for Al using the nugded elastic band method \citep{Esteban-Manzanares2019_CS}. $\nu_{eff}$ is an effective attempt frequency that is obtained from the product of  the fundamental attempt frequency (= 10$^{11}$ \citep{saroukhani2016harnessing, sobie2017modal}) with a scaling factor given by the ratio of the simulation strain rate (10$^4$ s$^{-1}$) to the experimental strain rate (10$^{-3}$ s$^{-1}$), following the strategy presented in \cite{Hussein2015}.  The scaling factor is used to have the same number of cross-slip attempts in the simulation and in the experiments in the same time interval. Finally, $\Delta H$ is the activation enthalpy for cross-slip that was computed from molecular dynamics simulations at different temperatures, leading to an analytical expression for the energy barrier as function of the Schmid stress on the cross-slip plane ($\sigma_S^{cs}$) and of the Escaig stresses on the glide ($\sigma_E^{g}$) and cross-slip ($\sigma_E^{cs}$) planes \citep{Esteban-Manzanares2019_CS}

\begin{equation}
\begin{split}
\Delta H(\sigma_E^g,\sigma_E^{cs},\sigma_S^{CS})=\Delta E_0-(V_E^g\sigma_E^g+V_E^{cs}\sigma_E^{cs}+V_S^{cs}\sigma_S^{cs})-\\ \frac{1}{2}[\Omega_1\sigma_E^g\sigma_E^{cs}+\Omega_2\sigma_E^g\sigma_S^{cs}+\Omega_3\sigma_E^{cs}\sigma_S^{cs}+\Omega_S^{cs}(\sigma_S^{cs})^2]
\end{split}
\label{Eq:Energy_barrier}
\end{equation}

\noindent where $\Delta E_0$ (= 0.582 eV) is the activation energy in the athermal limit, determined using the nugded elastic band method, while $V_E^g$, $V_E^{cs}$ and $V_S^{cs}$ stand for the activation volumes corresponding to the 
 Escaig stress on the glide and cross-slip planes  and to the Schmid stress on the cross-slip plane, respectively. Finally, $\Omega_1$, $\Omega_2$, $\Omega_3$ and $\Omega_S^{cs}$ stand for the polarization coefficients that account for the interaction of the applied stress with the local stress field created by the defect.  The activation volumes and polarization coefficients in eq. \eqref{Eq:Energy_barrier} can be found in \citep{Esteban-Manzanares2019_CS} and are depicted in tables \ref{Tab:CS_1} and \ref{Tab:CS_2} respectively. 

\begin{center}
\begin{table}[h]
\begin{center}
\begin{tabular}{ccc}
\toprule
 $V_E^g$ & $V_E^{CS}$ & $V_S^{CS}$\\
	\midrule
 13.6 & 11.6 & 9.4 \\
\bottomrule
\end{tabular}
\end{center}
\caption{Cross-slip activation volumes for Escaig stress on the glide ($V_E^g$) and cross-slip ($V_E^{cs}$) planes and for the Schmid stress on the cross-slip plane ($V_S^{cs}$) expressed in $b^3$, where $b$ is the Burgers vector of Al \citep{Esteban-Manzanares2019_CS}.}
\label{Tab:CS_1}
\end{table}
\end{center}

\begin{center}
\begin{table}[h]
\begin{center}
\begin{tabular}{cccc}
\toprule
 $\Omega_S^{cs}$ & $\Omega_1$ & $\Omega_2$ & $\Omega_3$\\
	\midrule
0.102 & -0.0516 & -0.0652 & -0.0294\\
\bottomrule
\end{tabular}
\end{center}
\caption{Cross-slip polarization coefficients expressed in $b^3$/MPa \citep{Esteban-Manzanares2019_CS}.}
\label{Tab:CS_2}
\end{table}
\end{center}

Finally, the term $(1-T/T_m)$ in eq. \eqref{eq:CS} takes into account the entropic contribution to the activation energy through the enthalpy-entropy Meyer-Neldel compensation rule \citep{Meyer1937}.

Cross-slip was introduced into the DDD simulations using a Metropolis MonteCarlo. To this end, the cross-slip probability defined in eq. \eqref{eq:CS} was evaluated for screw dislocation segments in the dislocation network  with a tolerance of  $\pm$ 2$^\circ$ between the direction of the Burgers vector and of the screw dislocation segment. Larger tolerances (up to 15$^\circ$ \citep{Hussein2015}) can be found in the literature but it was considered that larger tolerances  would change significantly the local stress state at the dislocation segments and influence artificially the cross-slip. Then, the Peach-Koehler force acting on the segment was projected to both the glide and the potential cross-slip planes and it was required that that projection of the Peach-Koehler force on the cross-slip plane was at least twice higher than that in initial glide plane to avoid numerical oscillations and to facilitate that the dislocation segments continued to glide in the cross-slip plane. Finally, the energy barrier and cross-slip probability were computed for the dislocation segment.  If the cross-slip probability was higher than 1, cross-slip occurred. If not, a random number in the range 0 to 1  was generated and cross-lip was allowed if the random number was lower than the cross-slip probability.

\section{Results}
\label{sec:Results}

\begin{figure}[t!]
\centering
\includegraphics[width=0.5\textwidth]{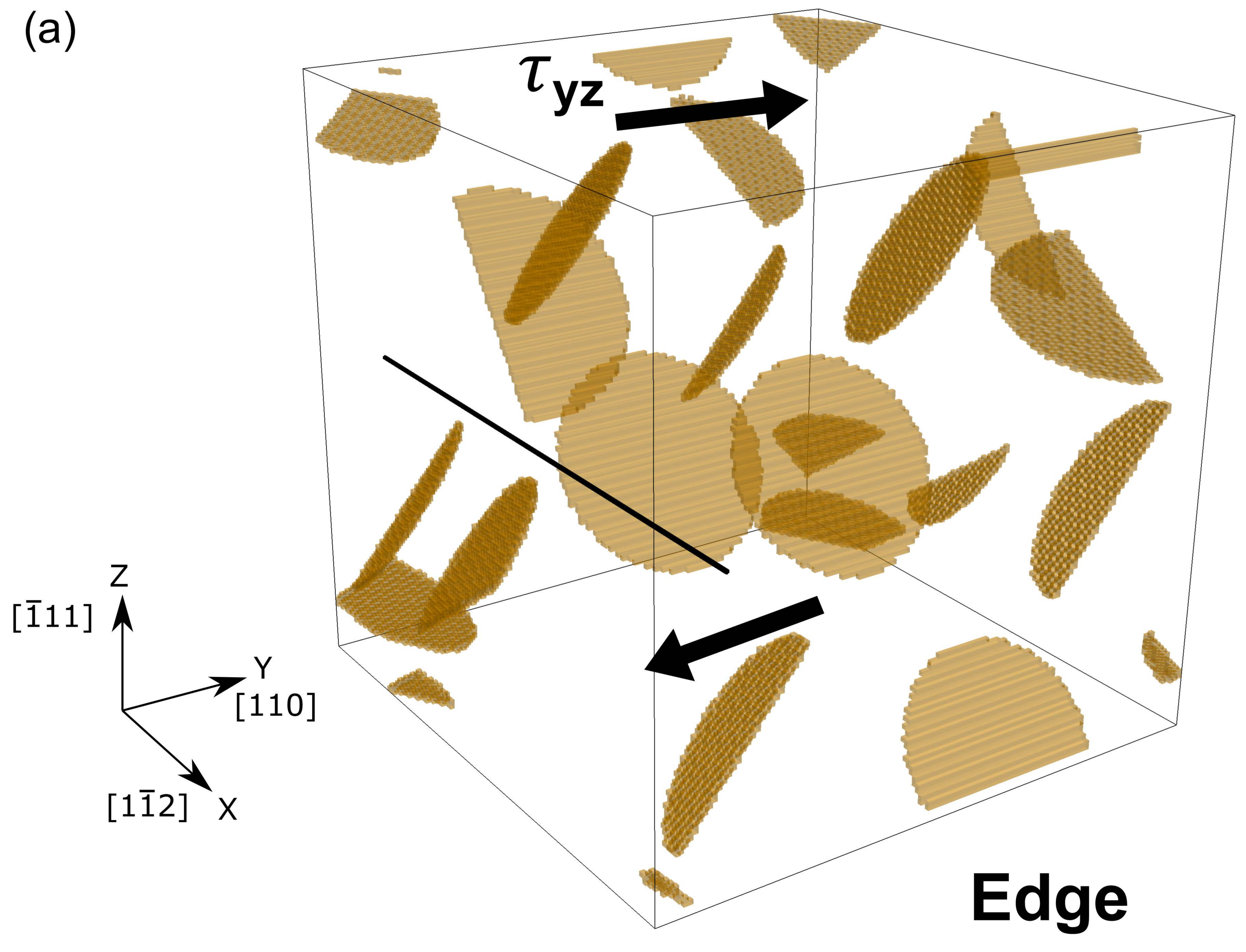} \vskip 5truemm
\includegraphics[width=0.5\textwidth]{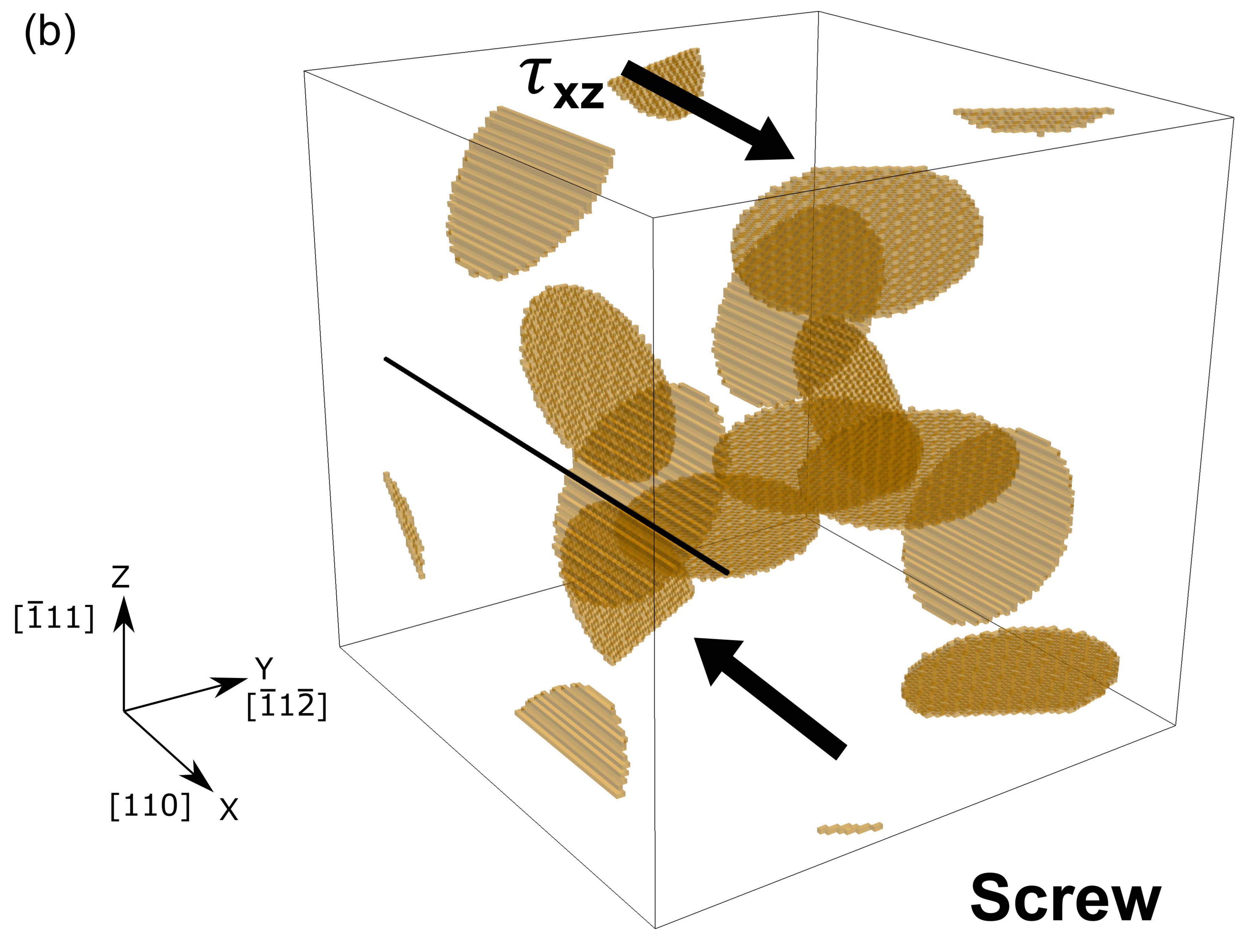}
\caption{Cubic simulation domain containing 12 $\theta'$ precipitates. (a) Initial edge dislocation. (b) Initial screw dislocation. Initial dislocations are represented by a straight black line.}  
\label{Distributions}
\end{figure}

The mechanical response of the Al-4 wt.\% Cu alloy was simulated using DDD in a cubic domain $V$ of  1 x 1 x 1 $\mu$m$^3$. The domain contained either a straight edge or screw dislocation, represented by a black line in Fig. \ref{Distributions}.  The axes of the cubic domain were parallel to the [1$\bar1$2], [110] and [$\bar1$11] orientations of the $\alpha$-Al lattice in the case of the ($\bar1$11)[110] edge dislocation (Fig. \ref{Distributions}a). They were parallel to the  [110], [$\bar1$1$\bar2$] and [$\bar1$11] orientations of the $\alpha$-Al lattice if the domain contained an ($\bar1$11)[110] screw dislocation (Fig. \ref{Distributions}b). Thus, the initial dislocation density in the simulations was 10$^{12}$ m$^{-2}$, which is reasonable for well-annealed crystals.

$\theta'$ precipitates were modelled as circular disks of 340 nm in diameter and 9 nm in thickness. 12 precipitates of these dimensions were introduced in the simulation box to attain a precipitate volume fraction of 1{\%}, in agreement with the experimental results summarized in Table \ref{tab:precipitate-size}.  Although the precipitate diameters in the alloy followed a log-normal distribution, all the precipitates in simulation box had the same dimensions. Sobie {\it et al.} \cite{SBC15} analyzed the strengthening due to voids and interstitial loops formed by irradiation by means of dislocation dynamics. The hardening from a normal distribution of defects was compared to that from the mean size, and was shown to have no statistically significant dependence on the distribution for both voids and loops. Precipitates of the three different orientation variants  (with the broad face parallel to the (100), (010) and (001) planes of the FCC $\alpha$-Al lattice) were included in the simulation domain. Each precipitate variant has 4 possible orientations of the SFTS due to the  4th fold rotational symmetry of the (001)$_\alpha$ planes and one of them was randomly assigned to each precipitate.

DDD simulations were carried out by applying a shear strain parallel to the ($\bar1$11) plane in the [110] direction at an applied strain rate of 10$^{4}$ s$^{-1}$ and a temperature $T$ = 300 K (Fig. \ref{Distributions}). In a previous investigation \citep{Santos-Guemes2018}, simulations at  strain rates in the range 10$^2$ to 10$^5$ s$^{-1}$  were carried out in similar domains containing only one precipitate. The results were compared  with those obtained using a relaxation strategy that allows to study the dislocation dynamics under quasi-static conditions. It was found that the stress-strain curves obtained at strain rates $\le$ 10$^{4}s^{-1}$ were equivalent to those found under quasi-static conditions. 

Equations of motion were integrated using an Euler explicit scheme in each time step of the DDD simulation. The dislocation velocity was determined from the Peach-Koehler force on the slip plane, according  to eq. \eqref{eq:mobility}.  In addition, the plastic strain accumulated in each time step was determined from the area swept by the dislocation line, as indicated in \citep{Bertin2018}. The effect of solid solution was included in the analysis through the friction stress $\tau_{ss}$ which depends on volume fraction of Cu atoms which remain in solid solution. Taking into account the precipitate volume fraction,  the remaining fraction of Cu atoms in solid solution was 3.5{\%} of Cu. Their contribution to the CRSS was quantified by \cite{Rodriguez-Veiga2018} and it was equal to 25 MPa.

The Peach-Koehler force acting on the dislocation line was determined in each time step by solving the mechanical equilibrium equations in the domain $V$ under periodic boundary conditions according to

\begin{equation} \label{MechEq}
 \left\{
 \begin{array}{l}
  \tensor{\sigma}(\mathbf{x})=\ftensor{C}(\mathbf{x}):[\tensor{\epsilon}(\mathbf{x})-\tensor{\epsilon}^p(\mathbf{x})-\delta(\mathbf{x})\tensor{\epsilon}^0_p], \quad \forall \mathbf{x} \in V
\\
 \mathrm{div} ( \tensor{\sigma}(\mathbf{x}))=0  \quad \mathbf{x} \in V \\
\tensor{\sigma}  \cdot \mathbf{n} \text{ opposite on opposite sides of } \partial V \\
\frac{1}{V}\int_V \tensor{\epsilon}(\mathbf{x})=E
 \end{array}
 \right.
\end{equation}

\noindent where $\ftensor{C}$ denotes the fourth order elasticity tensor (which is different in the matrix and in the precipitate), $\tensor{\epsilon}$ the total strain, $\tensor{\epsilon}^p$ the plastic strain, $\tensor{\epsilon}^0_p$ the SFTS (which is different in each precipitate depending on the orientation variant and on the orientation of the transformation strain  in the precipitate), $\partial V$  the boundaries of domain $V$ with normal $\mathbf{n}$ and $E$ the imposed macroscopic strain. $\delta(\mathbf{x})$ is the Dirac delta function which is equal to 1 inside the precipitates and equal to 0 in the Al matrix. The plastic strain $\tensor{\epsilon}^p(\mathbf{x})$ was computed directly from dislocation motion in the DCM \citep{Capolungo2015_DDDFFT}. The boundary value problem was analyzed using FFT solver \citep{Capolungo2015_DDDFFT,Bertin2018}  with a grid of 128 x 128 x 128 voxels. The details of the FFT algorithm can be found in \cite{Bertin2018}. It should be noted that the Gibbs fluctuations, associated with the FFT solver, due to the discontinuous strain field were attenuated by the use of rotational discrete gradient operators in Fourier space \cite{W15}.

12 cubic domains with different precipitate distributions were generated. Six included an initial edge dislocation and six an initial screw dislocation. The straight dislocation segments were introduced using the methodology presented in \cite{Santos-Guemes2018}. To this end, a rectangular prismatic loop parallel to one cube faces was introduced in the cubic domain and  two opposite sides of the loop  were moved in opposite directions until they reached the boundaries of the domain and annihilate each other leading to two straight dislocations forming a dipole within the domain. One of the dislocations was fixed during the simulation and the Field Dislocation Mechanics method was used to cancel the stress field created in the domain by the fixed dislocation following the methodology presented in \citep{Berbenni14,Brenner14,Djaka17}.

DDD simulations were carried out assuming that precipitates were impenetrable for dislocations but including different dislocation/precipitate interaction mechanisms to ascertain the role played for each one on the CRSS. For the sake of clarity, simulations without and with cross-slip will be discussed separately. 

\begin{figure}[h!]
\centering
\includegraphics[scale=0.8]{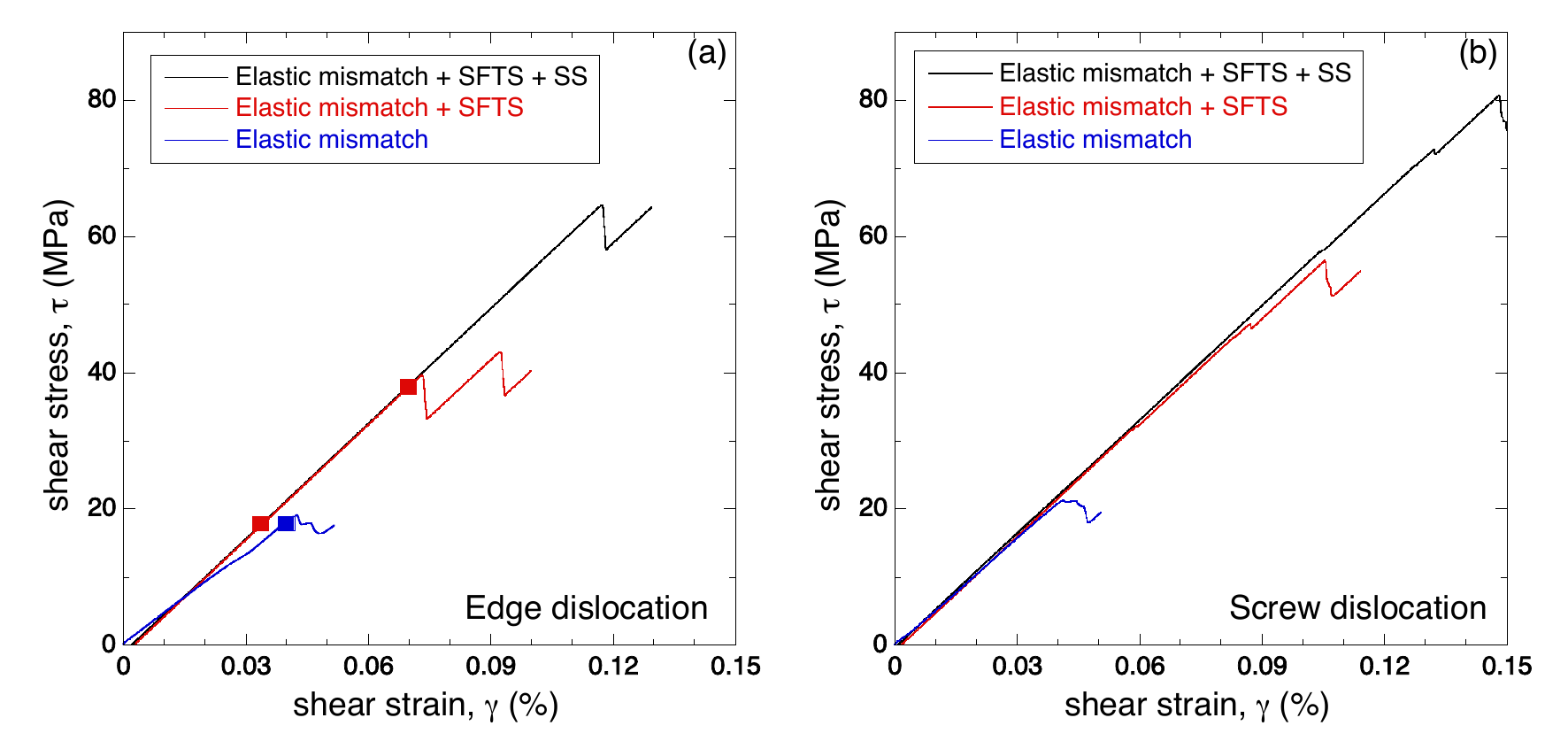}
\caption{Shear stress ($\tau$) - strain ($\gamma$) curves obtained from the DDD simulation of a cubic domain of the alloy. (a) Initial edge dislocation. (b) Initial screw dislocation. The blue and red squares correspond to the snapshots of the simulations in Figs. \ref{Snapshots-O} and \ref{snapshots-SFTS}, respectively.}  
\label{Strengthening}
\end{figure}

\subsection{Simulations without cross-slip}

The dislocation remains in the initial slip plane in the simulations without cross-slip and it is easy to monitor the progress of the dislocation depending on the physical dislocation/precipitate interaction mechanisms included in the analyses. In the first type of simulations, the elastic mismatch between the matrix and the precipitate was included in the analysis. The effect of the SFTS around the precipitates was added in the second set of simulations while the influence of solid solution strengthening was also included in the last set of simulations. The corresponding shear stress - shear strain curves are plotted in Figs. \ref{Strengthening}a and b for two simulation domains that initially contain an edge or a screw dislocation, respectively.  In all cases, the shear stress- strain curve are linear until the dislocation is able to overcome the precipitates and the initial CRSS is clearly defined by the first drop in the applied stress at the onset of non-linear deformation. 

\begin{figure}[t!]
\centering
\includegraphics[height=5.1cm]{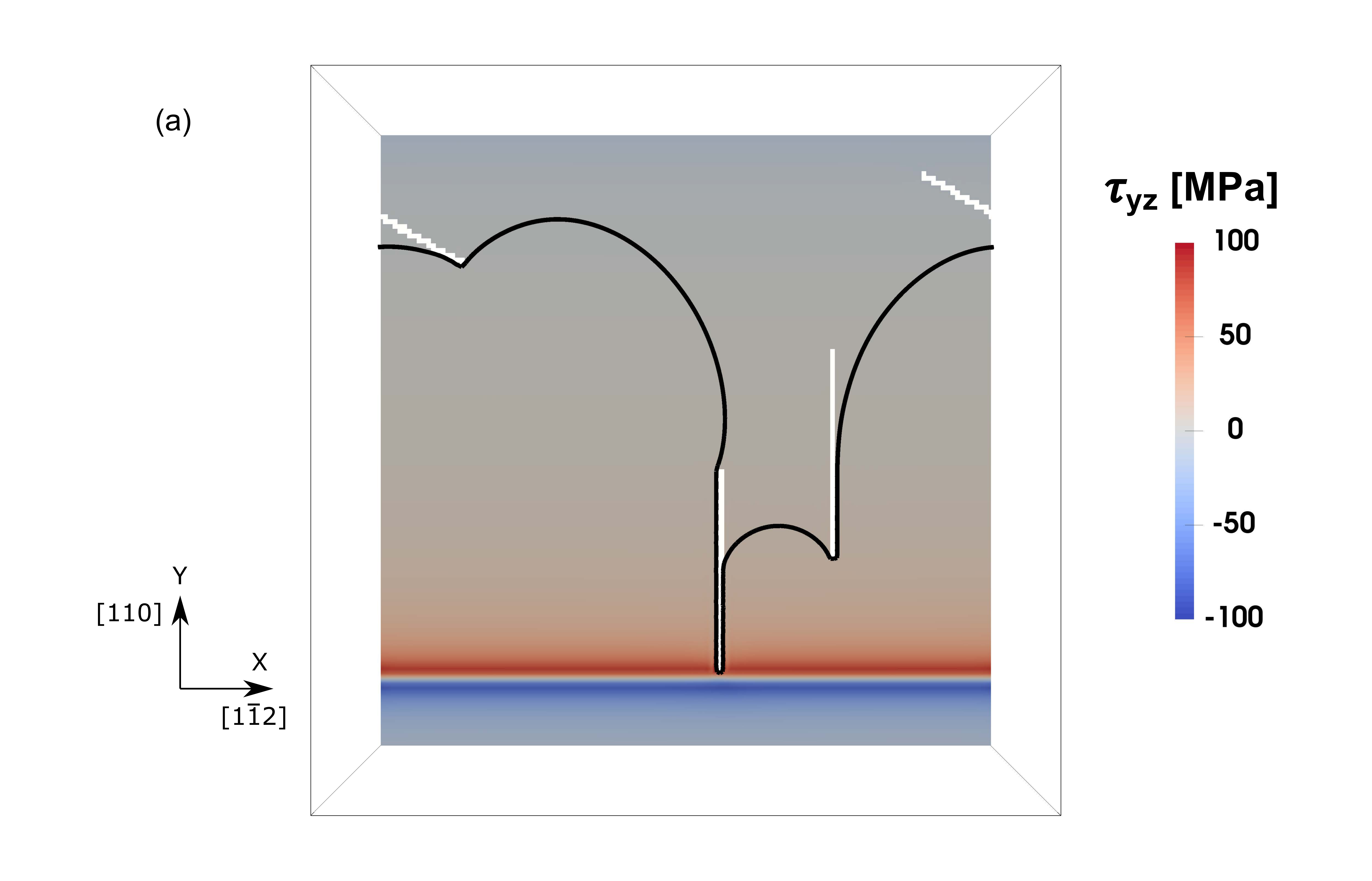}
\caption{Snapshot of the dislocation/precipitates interactions at the blue square in the shear stress-strain curve in Fig. \ref{Strengthening}a ($\tau$ = 18 MPa). The shear stress field in the image corresponds to the initial one induced by the edge dislocation.}  
\label{Snapshots-O}
\end{figure}

Large differences in the CRSS were observed for both edge and screw dislocations as a function of the interaction mechanisms included in the simulation. Simulations that only considered that the precipitates were impenetrable for dislocations and the image stress associated with the elastic mismatch led to CRSS slightly higher than 20 MPa. Moreover, the CRSS only decreased by $\approx$ 1 MPa if the elastic mismatch was not included in the simulations, in agreement with previous investigations \citep{Shin2003_DDD_FEM}. Under these circumstances, only the precipitates that intersect the glide plane are an obstacle to the dislocation glide and the CRSS is determined by the area and orientation of the cross-section of the precipitates in the glide plane. The evolution of the dislocation line in the slip plane is plotted together with the shear stress-strain curve in the movie found in the Supplementary Material (Movie S1, Dislocation-precipitate-Orowan.avi). One snapshot of this movie (that corresponds to the blue square in the shear stress-strain curves in Fig. \ref{Strengthening}a) is shown in Fig. \ref{Snapshots-O}. The applied shear stress was 18 MPa and the dislocation line was about to overcome the precipitates in the glide plane by making a loop between two precipitates (one perpendicular to the dislocation line and another located in the upper right of the glide plane). The  cross-section and relative location of these precipitates within the glide plane determined the CRSS.

\begin{figure}[t!]
\centering
\includegraphics[height=7cm]{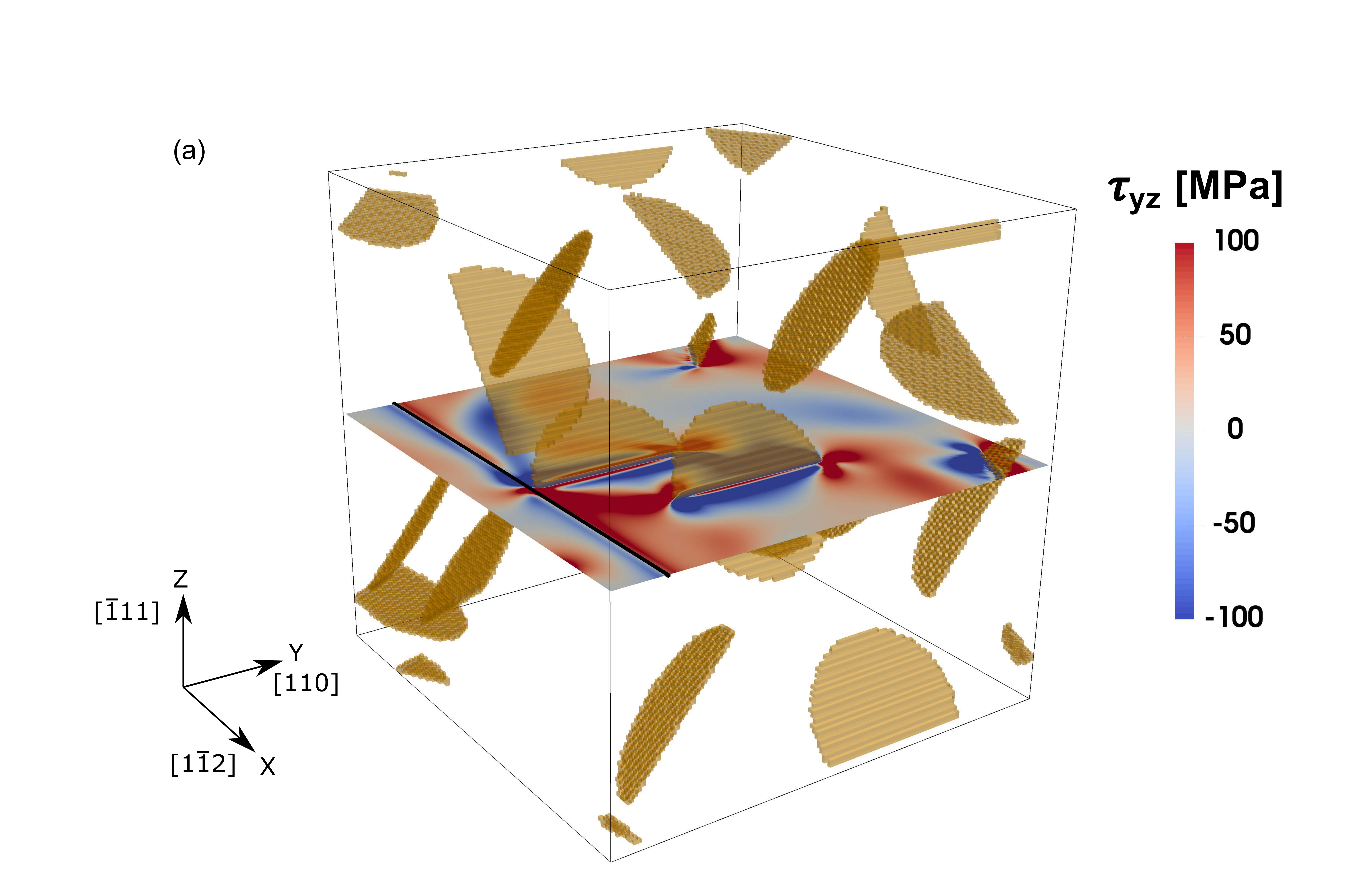} \vskip 5truemm
\includegraphics[height=5.1cm]{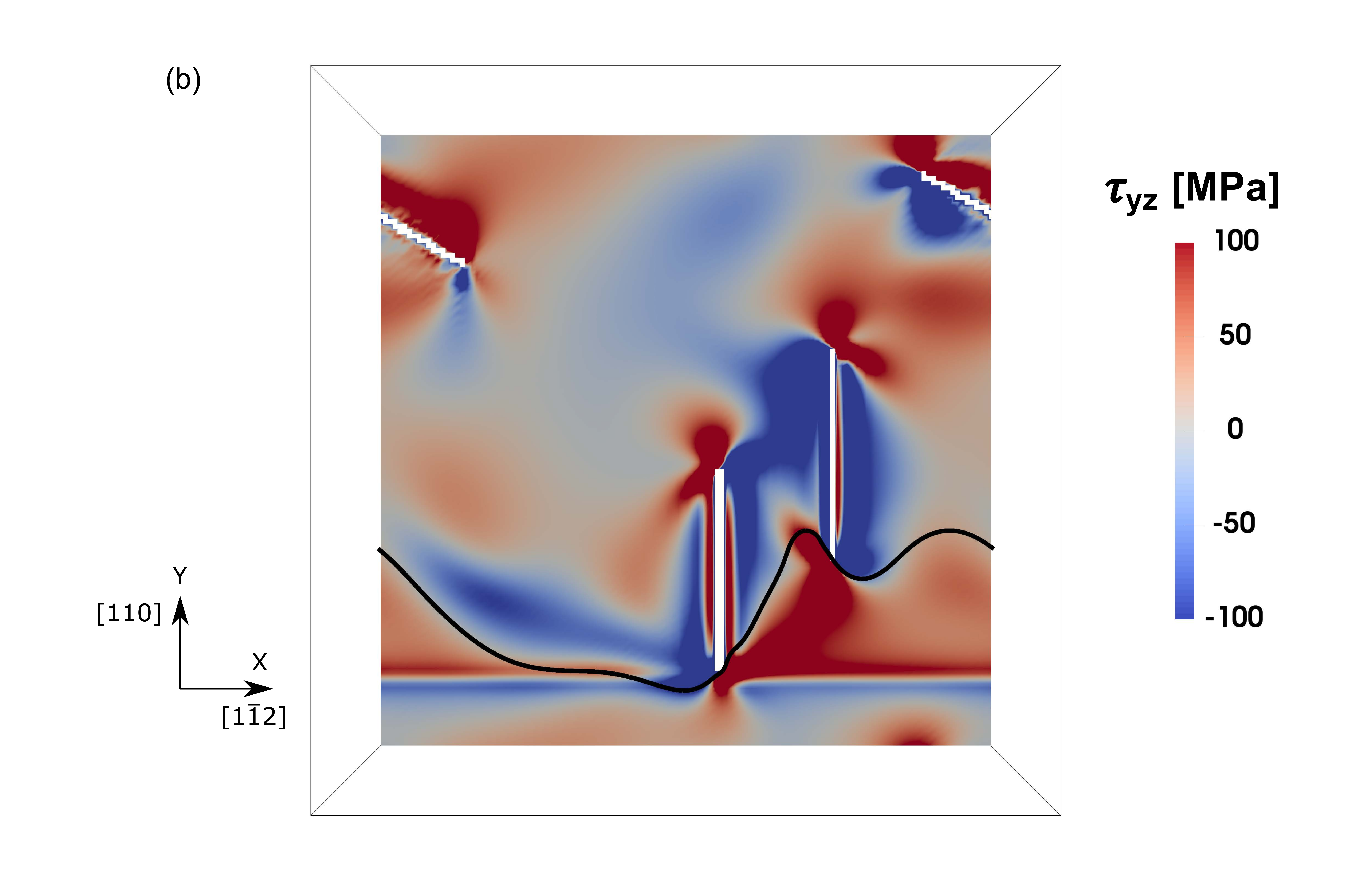}
\includegraphics[height=5.1cm]{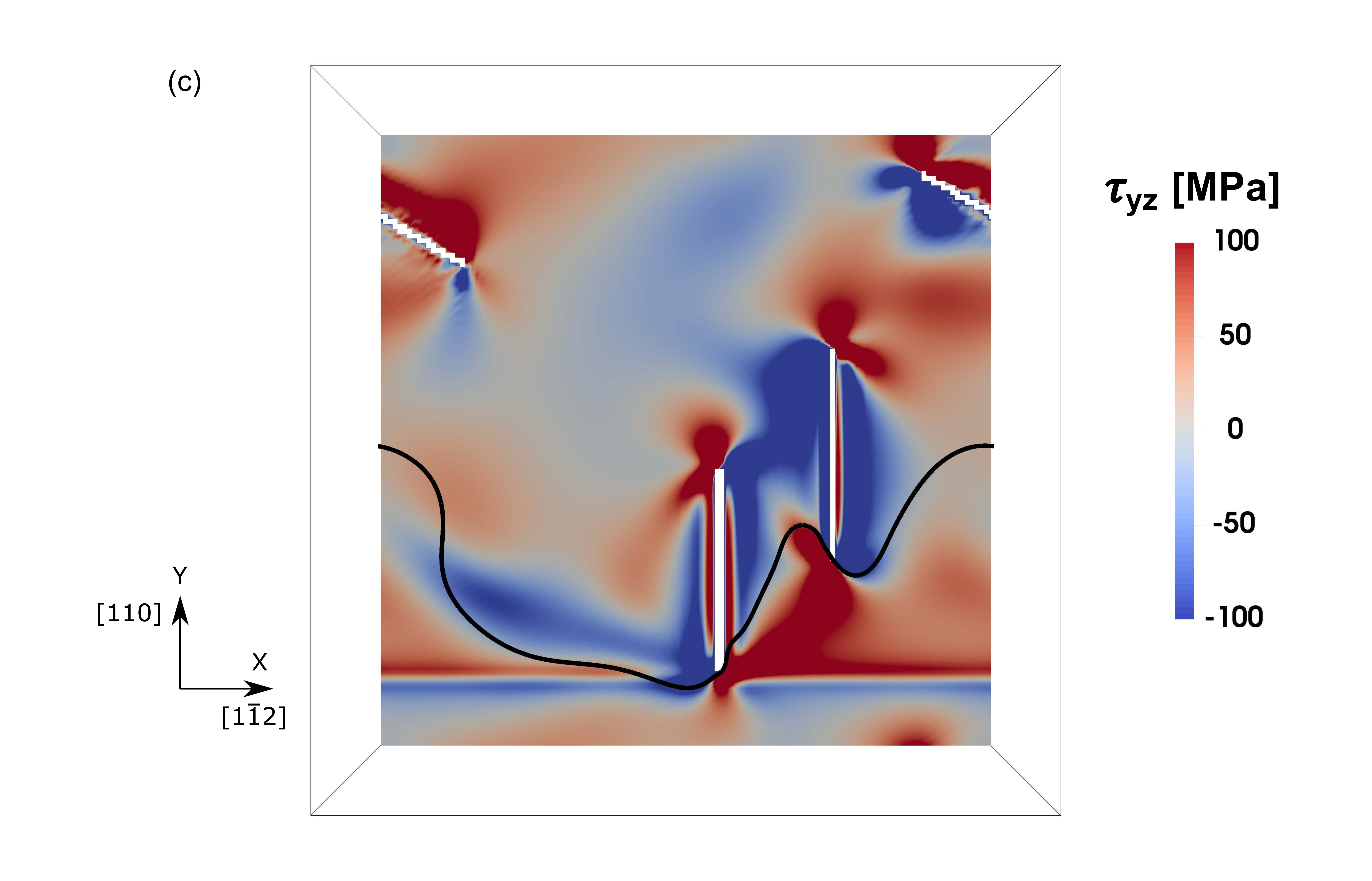}
\caption{(a) Shear stresses in the glide plane due to the contribution of the SFTS of the precipitates and of the dislocation line in a simulation domain containing 12 precipitates. The edge dislocation is shown as a black line. (b) Snapshot of the dislocation/precipitates interactions at the red square in the shear stress-strain curve in Fig. \ref{Strengthening}a ($\tau$ = 18 MPa). (c) {\it Idem} at the red square in the shear stress-strain curve in Fig. \ref{Strengthening}a ($\tau$ = 39 MPa). The shear stress field in the images always corresponds to the initial one induced by the edge dislocation and the SFTS of the precipitates.}  
\label{snapshots-SFTS}
\end{figure}

The CRSS increased by a factor of 2 or more if the SFTS around the precipitates were included in the simulations. The magnitude of these stresses is depicted in Fig. \ref{snapshots-SFTS}a which shows the shear stress in the glide plane due to the contribution of the SFTS and of the edge dislocation segment (black line) for a given simulation domain containing 12 precipitates. Negative (blue) values of the shear stress hinder the glide of the dislocation line, which is favoured to glide into regions with positive (red) shear stresses. The evolution of the dislocation line in the slip plane taking into account the effect of the SFTS is plotted together with the shear stress-strain curve in the movie found in the Supplementary Material (Movie S2, Dislocation-precipitate-Orowan-SFTS.avi). The influence of the SFTS is clearly observed in the snapshots of the movie shown in Fig. \ref{snapshots-SFTS}b and c that correspond to the red squares in the shear stress-strain curves in Fig. \ref{Strengthening}a when the applied shear stresses were 18 MPa and 39 MPa, respectively.  The dislocation line is trapped by the negative (blue) shear stress fields when the applied shear stress was 18 MPa (Fig. \ref{snapshots-SFTS}b) and it was necessary to increase the shear stress up to 39 MPa  (Fig. \ref{snapshots-SFTS}c) to by-pass the precipitate distribution. It is obvious from this figure that the main obstacle to the dislocation glide are not the precipitates but the large SFTS associated to them. Moreover, the most relevant SFTS correspond to precipitates that do not intersect the glide plane, indicating the 3D nature of the problem.

\begin{figure}[h!]
\centering
\includegraphics[width=0.8\textwidth]{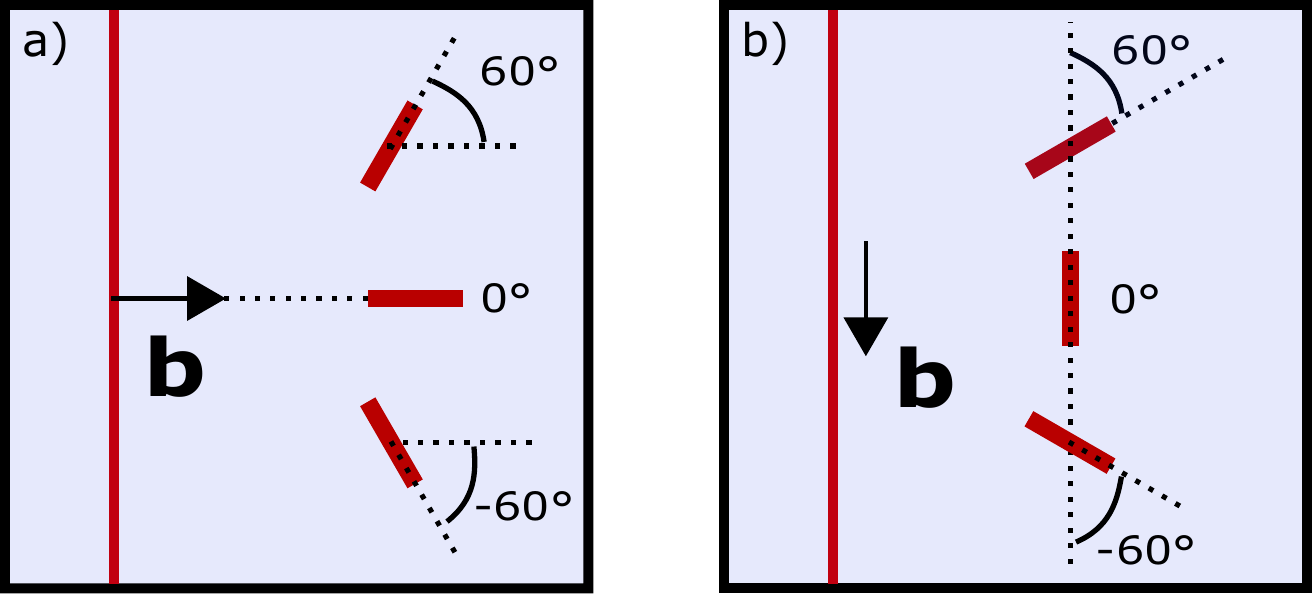}
\caption{ Schematic of the slip plane showing the relative orientation of the dislocation line and of the Burgers vector with respect to the intersection of the three precipitate variants with the slip plane. (a) Edge dislocation. (b) Screw dislocation.}  
\label{Orientation_precipitates}
\end{figure}

 It should be noticed that -on average- the CRSS obtained in the case of initial screw dislocations was higher than that in the case of initial edge dislocations. This difference can be explained considering the spatial distribution of the three precipitate orientation variants with respect to the Burgers vector of the dislocation. A schematic of the  slip plane showing the relative orientation of the dislocation line and of the Burgers vector with respect to the intersection of the three precipitate variants with the slip plane in the case of edge and screw dislocations is depicted in Figures \ref{Orientation_precipitates}(a) and (b), respectively. The movement of the screw dislocation (perpendicular to the Burgers vector) has to overcome one precipitate variant with the broad face parallel to the dislocation line, which is the strongest obstacle to the dislocation motion (Fig. \ref{Orientation_precipitates}(b)), while the precipitate variant by the broad face perpendicular to the dislocation line in Fig. \ref{Orientation_precipitates}(a) is the weakest obstacle for the propagation of the edge dislocation parallel to the Burgers vector. The precipitate variants oriented at +60$^\circ$ and -60$^\circ$ are in between those extreme cases.

Finally, the introduction in the simulations of the friction associated with the solution hardening led to a linear increase in the CRSS equal to the magnitude of $\tau_{ss}$ (Fig. \ref{Strengthening}). 

\subsection{Simulations with cross-slip}

Including cross-slip in the simulations led to a more complex dislocation/precipitate interaction pattern, as the dislocation cross-slips outside of the initial glide plane and interacts with precipitates, stress fields and other dislocations segments throughout the simulation box. The dislocation/precipitate interaction mechanisms in this case are depicted in the corresponding movie found in the Supplementary Material (Movie S3, Dislocation-precipitate-cross-slip.avi). This movie shows the progress of an initial edge dislocation through the simulation box from two different perspectives (the simulation box is observed from two perpendicular cubes faces). Dislocation segments in the initial slip plane are blue lines and change to red lines when screw segments cross-slip into another slip plane. Cross-slip occurs when the dislocations approach to the precipitates in certain orientations as a result of the interaction of the SFTS around the precipitates with the dislocations, leading to complex 3D dislocation patterns. 

\begin{figure}[h!]
\centering
\includegraphics[width=\textwidth]{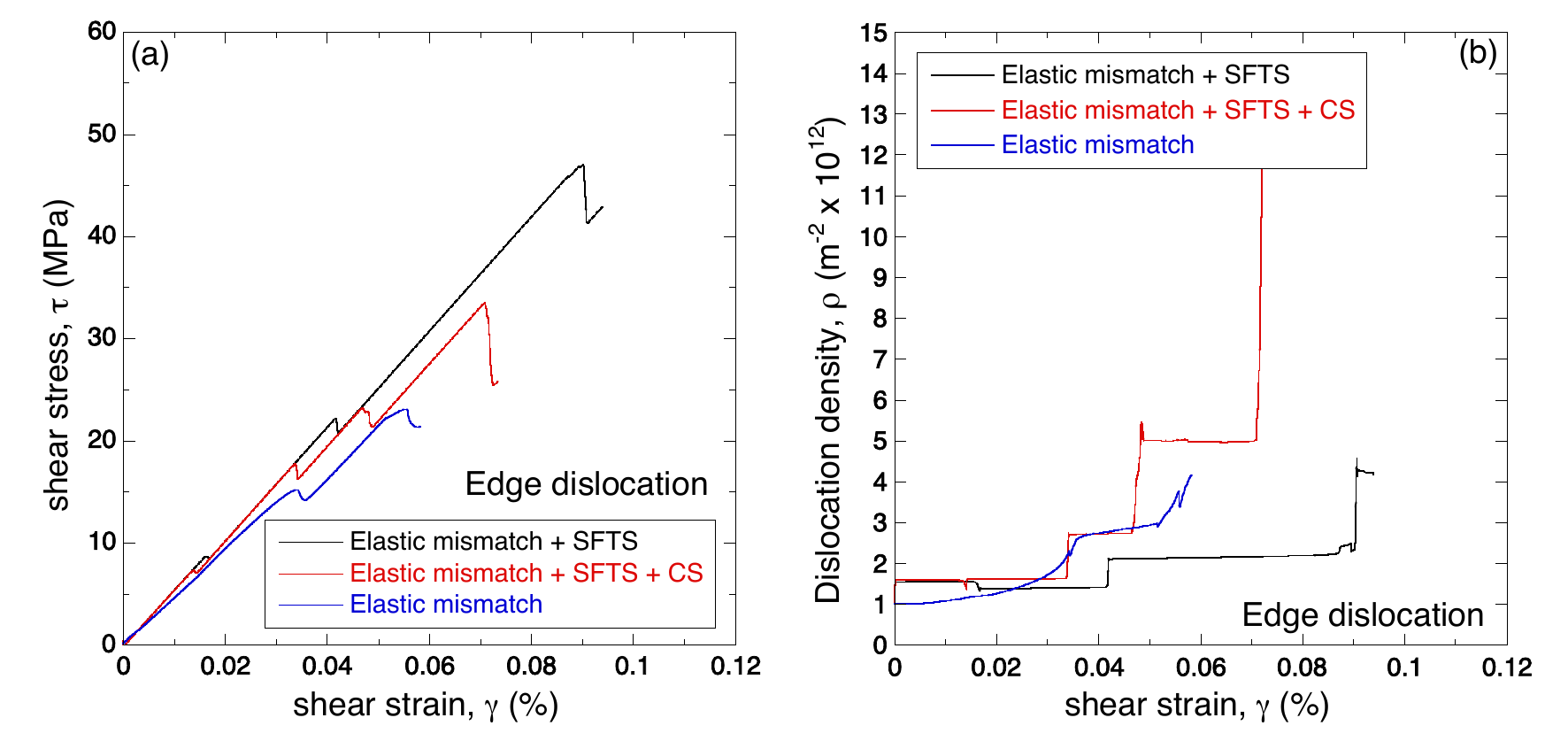}
\includegraphics[width=\textwidth]{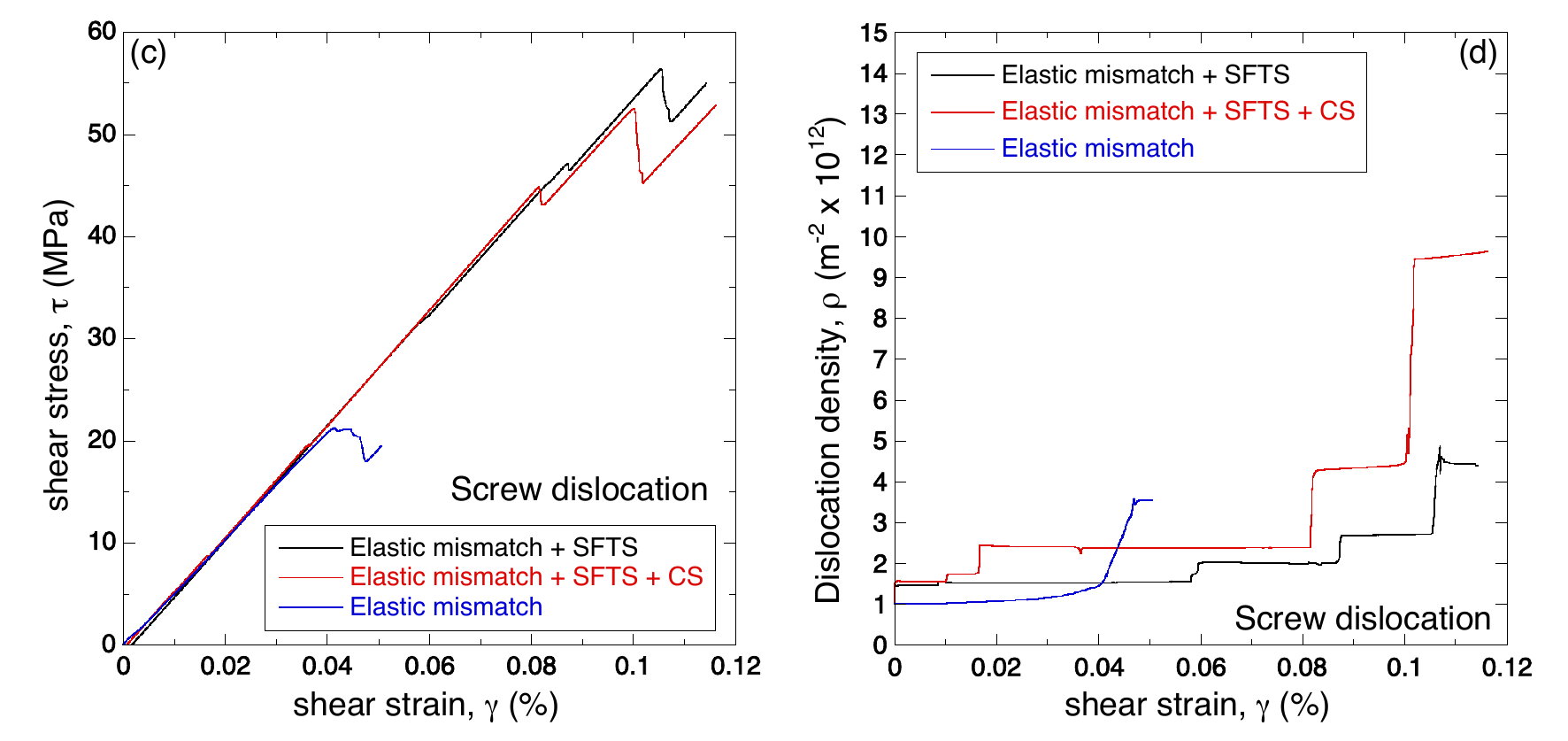}
\caption{(a) Shear stress-strain curves obtained from DDD simulations  of a cubic domain of the alloy containing an initial edge dislocation. (b) Dislocation density evolution for the DDD plotted in (a). (c) Shear stress-strain curves obtained from DDD simulations  of a cubic domain of the alloy containing an initial screw dislocation. (d) Dislocation density evolution for the DDD plotted in (c).}  
\label{CS}
\end{figure}

The presence of cross-slip changed the dislocation path through the precipitate forest looking for paths that presented lower resistance to the dislocation glide. This mechanism was more efficient in the case of precipitate distributions that presented a high resistance to the propagation of dislocations, as the one shown in  Fig. \ref{CS}a. This plot shows the shear stress-strain curves obtained from the DDD simulation of an initial edge dislocation assuming that the interaction of the dislocation with the impenetrable precipitates included different mechanisms (elastic mismatch, elastic mismatch plus SFTS and both mechanisms plus cross-slip). The effect of solid solution was not included because it only leads to a linear increase in the CRSS. The presence of the SFTS led to a large value of the CRSS (above 40 MPa that was the average value for an initial edge dislocation) but this CRSS decreased in the presence of cross-slip up to 30 MPa because cross-slip allowed the dislocation to find a lower resistance path. The evolution of the dislocation density for the three different dislocation/precipitate interaction mechanisms is plotted in Fig. \ref{CS}b. The CRSS is identified in Fig. \ref{CS}a by a sudden drop in the shear stress that corresponds to a sudden increase in the dislocation density $\rho$ as the dislocation overcomes the precipitates due to the storage of dislocations in the form of Orowan loops around the precipitates (Fig. \ref{CS}b). As expected, the dislocation density increased more rapidly in the presence of cross-slip but this higher dislocation density did not influence the hardening of the alloy in this initial stages of plastic deformation.

If the resistance of the precipitate realization to the propagation of the dislocation was similar to the average, the presence of cross-slip did not modify significantly the CRSS. This is shown in Fig. \ref{CS}c, which includes the shear stress-strain curves obtained from the DDD simulation of the propagation of an initial screw dislocation. The same mechanisms of dislocation/precipitate interaction indicated in Fig. \ref{CS}a are considered in this figure. Cross-slip only led to a slight reduction in the CRSS in this case, although the dislocation density was always higher in the simulations including cross-slip. As a result, cross-slip reduced slightly the average value of the CRSS and reduced more the differences in CRSS among different precipitate realizations.

\subsection{Comparison with experiments}

The CRSS was determined from the DDD simulations corresponding to 12 domains with different precipitate distributions (six with an initial edge dislocation and another six with an initial screw dislocation). Five different scenarios of dislocation/precipitate interactions were included in the simulation of each domain. Only the Orowan mechanism (precipitates are impenetrable by dislocations) was considered in the first case. The image stresses were added in the second case, while the effect of the SFTS was also included in the third case. The effect of solid solution strengthening was added to the other three mechanisms in the fourth scenario while dislocation cross-slip was finally included in the fifth case.

\begin{figure}[h!]
\centering
\includegraphics[scale=1.0]{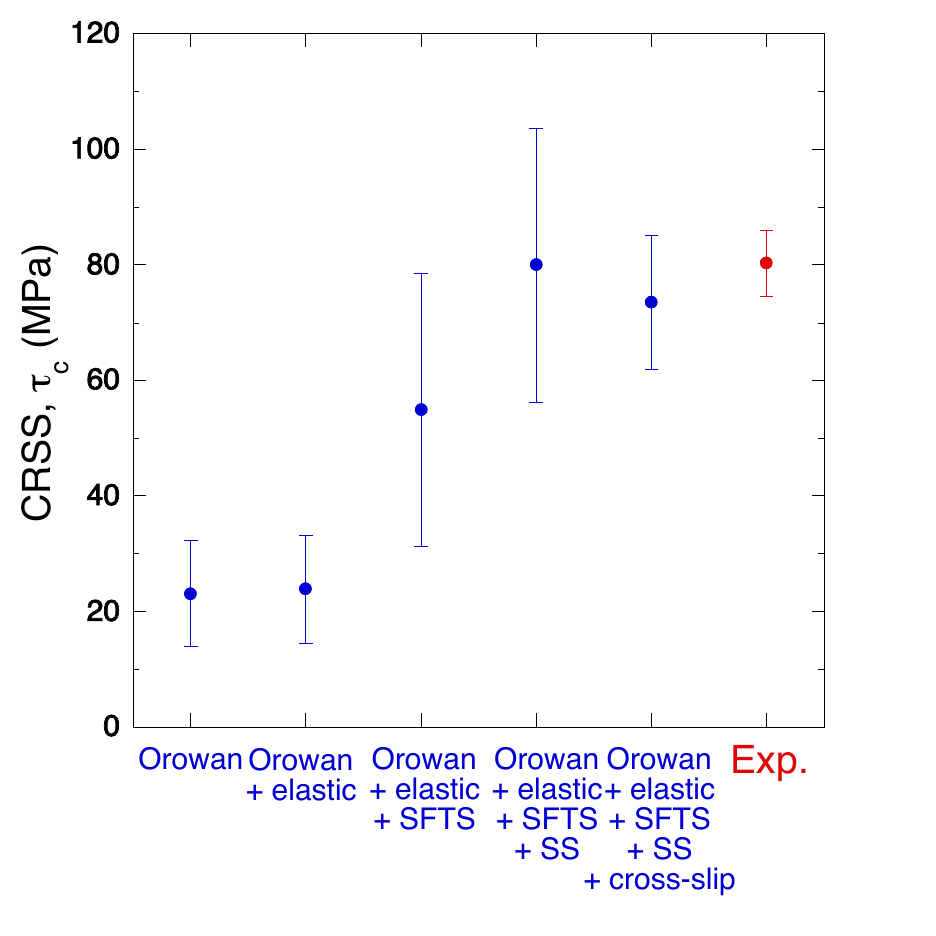}
\caption{DDD predictions of the CRSS for the Al-4wt.\%Cu aged at high temperature as function of the dislocation/precipitate interaction mechanisms and experimental results.}  
\label{Statistics}
\end{figure}

The average values of the CRSS corresponding to the 12 simulations for each scenario are plotted in Fig. \ref{Statistics} together with the corresponding standard deviations. The CRSS in the scenarios including only the Orowan mechanism or the Orowan mechanism plus the contribution of the images stresses due to elastic mismatch was in the range 15 to 30 MPa, far away from the experimental results (also included in Fig. \ref{Statistics}). The most relevant contributions to the precipitate strengthening in the case of the Al-Cu alloy containing $\theta'$ precipitates were induced by the SFTS and the solid solution of Cu atoms in the Al matrix. In particular, the SFTS increased the CRSS by $\approx$ 30 MPa due to the strong interaction of the dislocations with shear stresses induced by the transformation strain associated with the nucleation of the precipitate. In addition, the heterogeneity of the stress fields around the precipitates also led to a large increase in the scatter in the CRSS obtained with different precipitate distributions. This scatter was reduced when cross-slip was included in the simulations because the dislocations are allowed to propagate along different slip planes to overcome the barriers induced by the precipitates and by the SFTS. Nevertheless, cross-slip only reduced slightly the average value of the CRSS because it was not easy to overcome the obstacles induced by the precipitates and the associated stress fields even when cross-slip was allowed. 

Overall, the DDD predictions of the CRSS including all the physical mechanisms were in very good agreement with the experimental data and demonstrate the potential of DDD to make accurate predictions of precipitate-hardening in metallic alloys. The simulations also reveal why Al-Cu alloys present an outstanding response to precipitation hardening. The Orowan mechanism (which mainly depends on geometrical features and is present in all precipitation-hardened alloys) only contributes to $\approx$ 25\% of the CRSS and most of the hardening is provided by transformation stresses around the precipitate and by the large solid solution hardening induced by the Cu atoms \citep{LCH10}. Similarly, the contribution of solution hardening and of the coherency strains (due to the lattice misfit between the $\gamma$ matrix and the $\gamma'$ precipitates) to the overall strength is important in the case of precipitation-hardened Ni-based superalloys  \citep{GFM15, G18}: However, they do not contribute significantly to the CRSS of Mg alloys \citep{WLA19, HAM18}, that present a weaker response to precipitation hardening.  

It should be finally noted that that all the parameters that determine the interaction between the Al matrix and the precipitates were obtained from atomistic simulations or independent experimental observations and the results are free from adjustable parameters. In particular, lattice and elastic constants of the matrix and the precipitate were computed from DFT simulations \cite{Santos-Guemes2018}, dislocation mobility and cross-slip parameters were obtained from MD simulations \cite{Molinari2017_Mobility,Esteban-Manzanares2019_CS} and the solid solution contribution to the CRSS in addition to the details of the precipitate size, shape and volume fraction  were taken from experimental observations \cite{Rodriguez-Veiga2018}. Moreover, DDD simulations in combination with multiscale modelling strategies to simulate precipitate nucleation and growth during thermal treatments \cite{Liu2019} can be used to optimize precipitate hardening of metallic alloys or design new alloys improved mechanical properties \cite{BXL19}. In addition, the DDD strategy can also be used to analyze the effect of temperature on the CRSS for precipitation-hardened alloys (which enters in the simulations through the elastic constants of the phases and, more importantly, the cross-slip probability) as well as the initial work hardening rate due to the increase in dislocation density as a result of the interaction of dislocations with precipitates. 

\section{Conclusions}
\label{sec:Conclusions}

The CRSS for dislocation slip was determined in an Al-Cu alloy containing a homogeneous dispersion of the $\theta'$ precipitates using discrete dislocation dynamics. The precipitates were circular disks parallel to the \{100\} planes of the FCC Al lattice and the size, shape and volume fraction of the precipitates were obtained from TEM observations. The  precipitates were assumed to be impenetrable by the dislocations and the main parameters that determine the dislocation/precipitate interactions (elastic mismatch, stress-free transformation strains, dislocation mobility and cross-slip rate) were obtained from atomistic simulations. The discrete dislocation dynamics simulations were carried out within the framework of the discrete-continuous method which allows the inclusions of all the dislocation/precipitate interaction mechanisms and the mechanical fields during the analysis were obtained using a FFT solver.

The predictions of the CRSS were in agreement with the experimental results obtained by means of compression tests in micropillars oriented for single slip of the Al-Cu alloy. It should be noted that all the parameters in the simulations were obtained from either simulations or independent experiments, validating the predictive capability of the discrete dislocation dynamics simulations. In addition, the quantitative contribution of each mechanism to the CRSS could be ascertained from the simulations and provided an explanation for the strong response of Al-Cu alloys to precipitation hardening. It was found that the most important contribution to the CRSS  was caused by the interaction of the dislocations with the stress fields in the matrix due to the transformation strains associated with the formation of the $\theta'$ precipitates, followed by solid solution hardening and the Orowan contribution due to the bow out of the dislocations around the precipitates. The effect of the elastic mismatch between the matrix and the precipitates was negligible while the presence of cross-slip at ambient temperature reduced slightly the CRSS.

The bottom-up, multiscale simulation strategy presented in this paper opens the way to ascertain the different contributions to precipitate strengthening in metallic alloys and to optimize current alloys o design new ones based on the quantitative estimation of the influence of different factors (size, shape and spatial distribution of the precipitates, coherency and/or transformation strains, solid solution strengthening) on the CRSS. Moreover, the same simulation strategy can be used to analyze the effect of temperature on the CRSS and the initial work hardening rate due to the interaction between dislocations and dislocations and precipitates.

\section{Acknowledgments}

This investigation was supported by the European Research Council under the European Union's Horizon 2020 research and innovation programme (Advanced Grant VIRMETAL, grant agreement No. 669141). RS and BB  acknowledge the support from the Spanish Ministry of Education through the Fellowships FPU16/00770 and FPU15/00403.



\end{document}